\documentclass[11pt]{article}

\newcommand{\BE}{\begin{equation}}
\newcommand{\EE}{\end{equation}}
\newcommand{\BA}{\begin{eqnarray}}
\newcommand{\EA}{\end{eqnarray}}
\newcommand{\half}{{\scriptstyle{\frac{1}{2}}}}

\newcommand{\delv}{\vec{\nabla}}
\newcommand{\delsq}{\nabla^2}

\newcommand{\sech}{{\mbox{\rm sech}}}

\usepackage{amssymb}
\usepackage[dvips]{graphicx}

\begin{document}
\begin{titlepage}
\begin{flushright}
{\small DE-FG05-97ER41031-57 }
\end{flushright}
\vspace*{18mm}
\begin{center}
   {\LARGE{\bf Hydrodynamics of the Vacuum}}

\vspace*{15mm}

{\Large P. M. Stevenson}
\vspace*{3mm}\\
{\large T. W. Bonner Laboratory, Department of Physics and Astronomy\\
Rice University, P.O. Box 1892, Houston, TX 77251-1892, USA}
\vspace{18mm}\\
{\bf Abstract:}
\end{center}

Hydrodynamics is the appropriate ``effective theory'' for describing any 
fluid medium at sufficiently long length scales.  This paper treats 
the vacuum as such a medium and derives the corresponding hydrodynamic 
equations.  Unlike a normal medium the vacuum has no linear sound-wave 
regime; disturbances always ``propagate'' nonlinearly.  For an 
``empty vacuum'' the hydrodynamic equations are familiar ones (shallow 
water-wave equations) and they describe an experimentally observed phenomenon 
--- the spreading of a clump of zero-temperature atoms into empty space.  
The ``Higgs vacuum'' case is much stranger; pressure and energy density, 
and hence time and space, exchange roles.  The speed of sound is formally 
infinite, rather than zero as in the empty vacuum.  Higher-derivative 
corrections to the vacuum hydrodynamic equations are also considered.  
In the empty-vacuum case the corrections are of quantum origin and the 
post-hydrodynamic description corresponds to the Gross-Pitaevskii equation.  
I conjecture the form of the post-hydrodynamic corrections in the Higgs 
case.  In the $1 \! + \! 1$-dimensional case the equations possess 
remarkable `soliton' solutions and appear to constitute a new exactly 
integrable system.

\end{titlepage}

\newpage

%

\section{Introduction}

     Quite generally, any fluid medium when viewed at sufficiently long 
length scales can be described by hydrodynamics.  For length scales much, 
much greater than the mean free path the microscopic dynamics becomes 
irrelevant, except for determining the equation of state of the medium.  
The only relevant degrees of freedom are a density $n$ and a flow velocity 
$\vec{{\bf v}}$:  there is some `stuff' and it flows.  

     The idea of this paper is very simple:--- to apply hydrodynamics to the 
vacuum.  In fact, there are two cases; (i) an ``empty'' vacuum, where the 
equilibrium density is zero, and (ii) a ``spontaneous condensate,'' or 
``Higgs-type'' vacuum, where the relativistic energy density is minimized for 
some non-zero density $n=n_v$.  The actual physical vacuum is empty of 
certain particle species ({\it e.g.} electrons) but is a spontaneous 
condensate of others ({\it e.g.} quarks and gluons).  The electroweak 
theory requires the vacuum to have a nonzero vacuum expectation value for 
the Higgs field, which one can view as a spontaneous condensate of scalar 
particles.  This point has been well expressed by 't Hooft: ``What we 
experience as empty space is nothing but the configuration of the Higgs 
field that has the lowest possible energy.  If we move from field jargon 
to particle jargon, this means that empty space is actually filled with 
Higgs particles.  They have Bose condensed.'' \cite{tHooft,fntemech}

     The great virtue of a hydrodynamic approach is that it is essentially 
independent of the microscopic dynamics.  Hydrodynamics is almost as 
universal and fundamental as thermodynamics and it may yield important 
lessons for modern particle physics.  (This view is advocated from a 
different perspective in an important recent paper \cite{jnpp}.)  
In particular, hydrodynamics may perhaps provide a route to a deeper 
understanding of the ``Higgs-type'' vacuum, a phenomenon that is a vital, 
but experimentally untested feature of the Standard Model. 

      Hydrodynamics, of course, has great limitations.  It does not tell one 
how to produce or detect the excitations that it describes.   Nor does it even 
tell one the scale of the phenomena, which is governed by a parameter whose 
value is set by the microscopic dynamics.  I think it best to avoid 
speculations until the hydrodynamic equations and their solutions have been 
thoroughly explored.  That task is begun in this paper, but much more remains 
to be done.  For simplicity I concentrate on the $1 \! + \! 1$-dimensional 
case, though a spherically symmetric $d \! + \! 1$-dimensional solution will 
be discussed in Section 6.

%

\section{Hydrodynamics of a normal medium}

     In this subsection I very briefly review the basics of hydrodynamics 
applied to a ``normal,'' nonrelativistic medium \cite{landau,courant,chapman}. 
I assume that the medium is ``barotropic'' (energy density only a function of 
the pressure).  This is a good approximation for many normal media and will 
hold exactly in the vacuum case.  As a further simplification I concentrate 
on the $1 \! + \! 1$-dimensional case. 

     The hydrodynamic equations follow directly from conservation laws 
together with a constitutive relation for the medium, leading to coupled, 
first-order partial differential equations for $\rho$, the mass density, and 
$v$, the flow velocity.  Mass conservation and momentum conservation yield 
the fundamental equations (subscripts indicate partial derivatives: 
$\rho_t \equiv \partial \rho/\partial t$, etc.):
\BE
\label{mass}
\rho_t + (\rho v)_x = 0,
\EE
\BE
\label{mtm}
(\rho v)_t + (\rho v^2 + P(\rho))_x =0,
\EE
where $P(\rho)$ is the pressure as a function of density.  The previous 
equations together imply a simpler one, known as the Euler equation:
\BE
\label{Euler}
v_t + v v_x + \frac{P_x}{\rho} = 0.
\EE
Provided that the solution is everywhere smooth, one may regard Eqs. 
(\ref{Euler}) and (\ref{mass}) as the basic equations.  (However, if the 
solution develops discontinuities it is essential to remember that Eqs. 
(\ref{mass}) and (\ref{mtm}) are the fundamental pair.)

     For a normal medium the pressure varies linearly for small density 
disturbances:
\BE
P(\rho) = P_{\rm eq} + v_0^2 (\delta \rho) + O((\delta \rho)^2),
\EE
where $\delta \rho \equiv \rho -\rho_{\rm eq}$ and $v_0^2$ is a constant.  
For sufficiently small disturbances 
($\delta \rho \ll \rho_{\rm eq}$ and $v \ll v_0$) the hydrodynamic equations 
can then be linearized, yielding
\BE
v_t + v_0^2 \frac{(\delta \rho)_x}{\rho_{\rm eq}} \approx 0,
\EE
\BE
(\delta \rho)_t + \rho_{\rm eq} v_x \approx 0.
\EE
If the $x$ derivative of the first equation and the $t$ derivative 
of the second are combined, so as to eliminate $v$, one obtains the wave 
equation for $\delta \rho$:
\BE
(\delta \rho)_{tt} - v_0^2 (\delta \rho)_{xx} \approx 0.
\EE
(Similarly, one can derive the wave equation for $v$.)  Thus, in this 
linearized regime hydrodynamics reduces to acoustics.  Disturbances propagate 
as sound waves and obey the superposition principle.  Note that the speed of 
sound is determined by the thermodynamic derivative
\BE
v_0^2 = \left. \frac{\partial P}{\partial \rho} \right|_{\rm eq} .
\EE

    For larger disturbances, violating one or both of the conditions 
$\delta \rho \ll \rho_{\rm eq}$, and $v \ll v_0$, nonlinear effects come 
into play.  In the vacuum case, as we shall see, there is no linear, acoustic 
regime; disturbances are always nonlinear.  Insight into the nonlinear 
regime can be gained by considering first the case of a pressureless fluid, 
where the Euler equation reduces to Burgers' equation (also known by many 
other names)
\BE
\label{Burgers}
v_t + v v_x =0.
\EE
The solution, easily verified by direct differentiation, can be expressed 
as follows \cite{jnppnote}:  Let the initial condition be $v(x, t=0)=V(x)$, 
then at later times 
\BE
v=V(\chi) \quad \quad {\mbox{\rm with}} \quad \quad \chi=x-V(\chi) t,
\EE
where the latter equation defines $\chi$ implicitly.  Thus, velocity 
disturbances propagate locally at the local flow velocity.  
If $V$ is monotonic increasing to the right then the forward 
part of the wave outruns the backward part and the disturbance simply 
stretches out.  However, if $V$ is monotonic decreasing then the backward part 
of the wave tends to catch up with the forward part, steepening the wave 
profile, as with ocean waves approaching the shore.  After a finite time 
the solution, rather than becoming multivalued, as with waves at a beach, 
develops a finite discontinuity in $v$, a shock wave.

   Quite generally, the solution may develop shocks (finite discontinuities 
in $\rho$ and/or $v$) in finite time, even from very smooth initial 
conditions.   Requiring mass and momentum conservation yields two equations:
\BE
\label{shock}
{\mbox{{\rm shock speed}}} = \frac{[[\rho v]]}{[[\rho]]} 
= \frac{[[\rho v^2 + P(\rho)]]}{[[\rho v]]},
\EE
where $[[X]] \equiv X_{\rm right} - X_{\rm left}$. The shock solution, 
even though the first derivatives do not exist at the shock, can be regarded 
as a ``weak solution,'' a solution in a distribution-theoretic sense, of the 
original equations \cite{evans}.  An important warning here is that the 
mathematical notion of a ``weak solution'' must be applied to the fundamental 
mass and momentum conservation equations, (\ref{mass}) and (\ref{mtm}), and 
{\it not} to the Euler equation (\ref{Euler}). 

    At a shock the solution violates the basic hydrodynamic assumption 
that all relevant length scales are much greater than the mean free 
path.  Thus, corrections to hydrodynamics play a role.  These corrections 
can be developed as a series expansion in (mean free path)/(length scale), 
known as the Chapman-Enskog expansion \cite{chapensk}.  For a normal medium 
the post-hydrodynamic corrections involve dissipative effects such as 
viscosity.  The correction terms involve second derivatives and violate 
(macroscopic) time-reversal invariance, since they entail entropy generation.  
In the pressureless-fluid case one encounters the viscous Burgers' equation 
\cite{fnteHopf}.
\BE
\label{vBurgers}
 v_t + v v_x = \epsilon v_{xx},
\EE
in which the viscosity term $\epsilon v_{xx}$ smooths out the shock 
discontinuity.  For this equation, and more generally, one may find the 
shock profile by looking for ``travelling wave'' solutions with $v=v(x-a t)$ 
and $\rho=\rho(x-a t)$ where $a=$ is a constant and $v_L \! \neq \! v_R$ 
and/or $\rho_L \! \neq \! \rho_R$, where 
$v_L \! \equiv \! v({\scriptstyle{x \to -\infty}})$, etc..  
As $\epsilon \to 0$ the travelling wave's profile narrows, approaching 
a discontinuity, and the speed $a$ tends to the shock speed of 
Eq.~(\ref{shock}).  In this sense, a hydrodynamic solution with shocks is 
still a meaningful approximate description of the physics.

%

\section{Constitutive relation of the vacuum medium}

     Henceforth a relativistic framework will be adopted; energy and energy 
density will include the rest-energy ($mc^2$) contribution and the speed of 
light, $c$, will be set to 1.  

      The ``empty vacuum'' and ``Higgs vacuum'' cases are illustrated in 
Figs. 1 and 2.  In both cases, since we are dealing with a zero-temperature 
medium, the pressure is given by 
\BE
\label{pressure}
P = -{\cal E} + n \frac{d {\cal E}}{d n},
\EE
where ${\cal E}$ is the energy density and $n$ is the number density of 
particles.  Note that a precise definition of $n$ is not needed.  What 
matters physically is the ``constitutive relation'' --- how $P$ varies with 
${\cal E}$ in the vicinity of the equilibrium state, and it is merely a 
convenience to express this relation parametrically as 
${\cal E} = {\cal E}(n)$ and $P=P(n)$ via the above equation.  
If, for example, the underlying microscopic theory is a scalar quantum field 
theory, one may take $n$ to be proportional to $\phi^2$, where 
$\phi$ is the classical field, and identify ${\cal E}(n)$ with the 
field-theoretic effective potential $V_{\rm eff}(\phi)$. 

\begin{figure}[htp]
\begin{center}
\includegraphics[width=7cm]{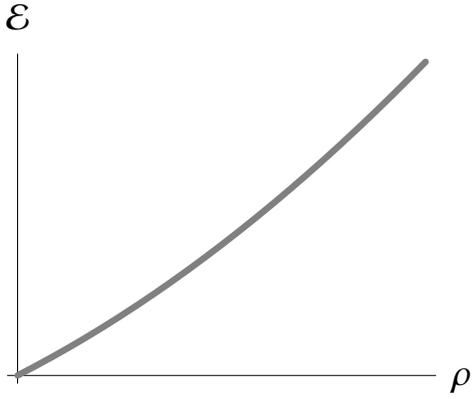}
\caption{The empty vacuum case; energy density as a function of 
mass density $\rho = m n$. The equilibrium density is $\rho=0$.}
\end{center}
\end{figure}

\begin{figure}[htp]
\begin{center}
\includegraphics[width=8cm]{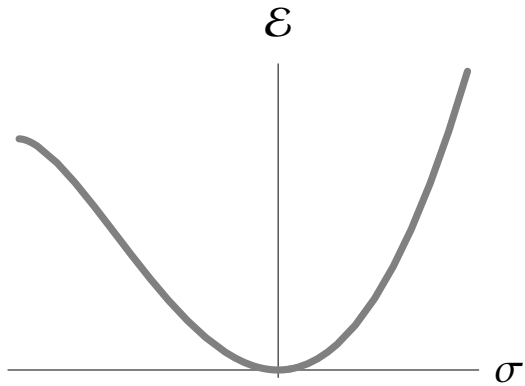}
\caption{The Higgs vacuum case; energy density as a function of 
$\sigma = \frac{1}{C} \left(\frac{n}{n_v}-1 \right)$ where ${\cal C}$ 
is a constant. }
\end{center}
\end{figure}

The empty-vacuum case is essentially nonrelativistic.  The energy density 
${\cal E}$ is dominated by the mass density (times $c^2$, but $c=1$ here).  
However, this linear term exactly cancels in the pressure (\ref{pressure}), 
so that the small non-linear term $\half g \rho^2$ is crucial \cite{fnteg}:
\BE
\label{empty1}
{\cal E} = \rho + \half g \rho^2 + \ldots, 
\EE
\BE
\label{empty2}
P =  \half g \rho^2 + \ldots .
\EE
The speed of sound is formally given by
\BE
v_0^2 \equiv 
\left. \frac{\partial P}{\partial {\cal E}} \right|_{\rho=0} = 
\left. \frac{d P}{d \rho} \right|_{\rho=0} = 0.
\EE

In the Higgs-vacuum case we find a ``mirror-image'' situation with the forms 
of ${\cal E}$ and $P$ interchanged.  (For convenience I subtract a constant 
from the energy density so that ${\cal E}$ vanishes in equilibrium. 
Since gravity is not being included in the discussion this step is quite 
innocuous.)
\BE
\label{Higgs1}
{\cal E} =  \half C \sigma^2 + \ldots ,
\EE
\BE
\label{Higgs2}
P = \sigma + \half C \sigma^2 + \ldots,
\EE
with \cite{fnteneg} 
\BE
\sigma = \frac{1}{C} \left(\frac{n}{n_v}-1 \right) ,
\EE
where $C$ is a positive constant.  $C$ is the compressibility of the vacuum 
medium
\BE
C = \left( \left. 
n_v^2  \frac{d^2 {\cal E}}{d n^2} \right|_{n=n_v} \right)^{-1}.
\EE
For small disturbances one has $P \gg {\cal E}$, so the Higgs-vacuum case is 
ultrarelativistic.  The speed of sound is formally infinite \cite{cpt01}:
\BE
v_0^2 \equiv 
\left. \frac{\partial P}{\partial {\cal E}} \right|_{\sigma=0} = 
\left. \frac{d P/d \sigma}{d {\cal E}/d\sigma} \right|_{\sigma=0} = 
\frac{1}{0} = \infty .
\EE
This is the first sign that something very interesting --- not to say 
bizarre --- is going on.  Note that the vanishing denominator arises because 
$d{\cal E}/dn$ vanishes at $n_v$, which is the defining property of a 
Higgs-type vacuum.  

     To anticipate (and oversimplify), the empty-vacuum case behaves like 
Burgers equation $v_t+v v_x=0$ and effects tend to ``propagate'' at the 
local flow velocity $v$, while the Higgs-vacuum case behaves like the 
$x \leftrightarrow t$ Burgers equation $v_t + \frac{1}{v} v_x =0$ whose 
implicit solution $v=v(x-\frac{1}{v}t)$ implies that disturbances tend to 
``propagate'' at speed $\frac{1}{v}$ (that is $c^2/v$, restoring $c$).  
Thus, the ``propagation'' speed is superluminal, and can be arbitrarily large 
as $v$ becomes arbitrarily small.  This is a fundamental and inescapable 
characteristic of the Higgs-type vacuum.  Obviously, deep issues related 
to causality are involved.  Note that it is well known that an 
ultrarelativistic medium (where the pressure is much greater than the 
energy density) can have a superluminal speed of sound \cite{URmedium1}.  
Discussions of the causality issue for an ultrarelativistic medium can be 
found in Ref. \cite{URmedium2,URmedium3}.  I set this issue aside for the 
present.

%

\section{Derivation of the vacuum hydrodynamic equations}

  The energy-momentum tensor for a perfect fluid has the form 
[$g^{\mu\nu}= {\rm diag}(1,-1,-1,-1)$]
\BE
      T^{\mu \nu} = ({\cal E} + P)u^\mu u^\nu - P g^{\mu \nu},
\EE
where ${\cal E}$ and $P$ are respectively the energy density and the pressure 
in the co-moving frame, and $u^\mu$ is the flow 4-velocity 
\BE
       u^\mu = (u_0,\vec{u}) = (\gamma, \gamma \vec{{\bf v}}),
\EE
with $\gamma \equiv 1/\sqrt{1-v^2}$. 
Energy and momentum conservation equations follow from 
$\partial_\mu T^{\mu \nu}=0$ with, respectively, $\nu=0$ and $\nu=i=1,2,3$: 
\BE
\label{em1}
\frac{\partial}{\partial t}\left[ ({\cal E} + P)u_0^2 -P \right] + 
\frac{\partial}{\partial x^j}\left[ ({\cal E} + P)u^j u_0 \right] = 0,
\EE
\BE
\label{em2}
\frac{\partial}{\partial t}\left[ ({\cal E} + P)u_0u^i \right] + 
\frac{\partial}{\partial x^j}\left[ ({\cal E} + P)u^j u^i + P \delta^{ij} 
\right] = 0.
\EE
Substituting for ${\cal E}$ and $P$ --- from either 
(\ref{empty1}, \ref{empty2}) in the empty-vacuum case, or 
(\ref{Higgs1}, \ref{Higgs2}) in the Higgs-vacuum case --- yields the 
fundamental equations of ``vacuum hydrodynamics.''  
  
     Provided the solution is free of shock discontinuities and singularities 
such as vortex lines, one may work with equivalent equations that are somewhat 
neater.  (Cf. the discussion below Eq. (\ref{Euler}).)  These neater equations 
arise from $u_\nu \partial_\mu T^{\mu \nu}=0$ (``energy conservation in the 
co-moving frame'') and 
$(g_{\alpha \nu} - u_\alpha u_\nu) \partial_\mu T^{\mu \nu} =0$ (``Euler 
equation'') \cite{degroot}.  In the empty vacuum case (in $1 \! + \! 1$ 
dimensions) these equations are:
\BE
\label{rel1}
(\gamma \rho)_t + (\gamma \rho v)_x =0,
\EE
\BE
\label{rel2}
((1+g \rho) \gamma v)_t + ((1+ g \rho) \gamma)_x =0.
\EE
Note that one could re-scale $\rho$ to 
$\hat{\rho} = g \rho$ and hence eliminate $g$ from the equations.

   Since one is interested in small disturbances it is natural to assume 
that $v$ and $g \rho$ are small.  I shall refer to this as the {\it 
nonrelativistic-flow approximation} (NFA).  In the empty-vacuum case it is 
equivalent to nonrelativistic reduction.  Restoring the factors of $c$ and 
considering $1/c$ to be a small parameter, $\epsilon$, one obtains the 
bookkeeping rules for the NFA \cite{fnteslow}
\BE
\label{NFAev}
v=O(\epsilon), \quad\quad g \rho = O(\epsilon^2), \quad\quad 
\frac{\partial}{\partial t} = O(\epsilon).
\EE
Applying these 
rules to the above equations yields
\BE
\label{ev1}
\rho_t + (\rho v)_x =0,
\EE
\BE
\label{ev2}
v_t + \left( \half v^2 + g \rho \right)_x =0.
\EE
Comparing these equations with (\ref{mass}), (\ref{Euler}), one sees that they 
correspond to an otherwise ``normal'' medium with a pressure function
$P(\rho)=\half g \rho^2$.   This is a special case of a ``polytropic gas'' 
($P \propto \rho^n$) with polytropic index $n=2$ 
\cite{landau,courant,chapman,fntelax}.  Ordinary gases are, to a 
good approximation, polytropic, but with $n$ between $1$ and $5/3$.  
However, the $n=2$ case arises in the treatment of shallow water waves 
(wavelengths long compared to the depth of the water), where $\rho$ is 
proportional to the vertical displacement of the water surface \cite{courant}.

    More relevantly, these equations also arise in the description of the
free expansion of an atomic Bose-Einstein condensate when the trapping 
potential is turned off \cite{stringari,dalfovo}.  This fact is natural from 
the present point of view: the clump of atoms, localized with a particular 
density distribution at $t=0$, is just a particular disturbance of the empty 
vacuum.  Provided that the initial density distribution is sufficiently 
slowly varying, its spreading will naturally be governed by the vacuum 
hydrodynamic equations.  This important example is discussed in more detail 
in Section 6. 

     There is an important conceptual subtlety with the NFA.  The original 
equations are Lorentz invariant, while the NFA equations are only Galilean 
invariant ({\it i.e.}, only approximately Lorentz invariant for small boosts). 
Thus, using the NFA implies restricting oneself to a particular Lorentz frame 
and its ``neighbouring'' frames.  For a normal medium the rest frame of the 
medium obviously plays this role.  However, the vacuum is a Lorentz-invariant 
medium; it has no rest frame.  The appropriate frame for the NFA is 
determined solely by the initial conditions.  If in some frame the NFA 
conditions are satisfied at $t=0$ then they will remain satisfied at all 
later times.  One may trivially take a NFA solution and boost it by a 
large Lorentz boost to obtain an approximate solution to the original 
relativistic equations in which $v$ is everywhere close to 1.  Only when the 
{\it range} of $v$ values is a significant fraction of unity is it necessary 
to abandon the NFA and return to the relativistic equations, (\ref{rel1}, 
\ref{rel2}).    

      In the Higgs vacuum case the pressure and the energy density exchange 
roles.  As a consequence, time and space in some sense exchange roles.  
In the $1 \! + \! 1$ dimensional case one obtains
\BE
(\gamma \sigma)_x + (\gamma \sigma v)_t =0,
\EE
\BE
((1+C \sigma) \gamma v)_x + ((1+ C \sigma) \gamma)_t =0,
\EE
which are the same as the empty vacuum case with $x \leftrightarrow t$ (and 
$\rho \to \sigma$, $g \to C$).  (In higher dimensions the 
``space $\leftrightarrow$ time'' interchange is obviously more complicated.  
See Eqs. (\ref{hvd1}, \ref{hvd2}) below.)

     Using this symmetry one can define a consistent NFA by the bookkeeping 
rules: 
\BE
\label{NFAhv}
v=O(\epsilon), \quad\quad C \sigma = O(\epsilon^2), \quad\quad 
\frac{\partial}{\partial x} = O(\epsilon).
\EE
and obtain the equations
\BE
\label{hv1}
\sigma_x + (\sigma v)_t =0,
\EE
\BE
\label{hv2}
v_x + \left( \half v^2 + C \sigma \right)_t =0.
\EE
Although the flow velocity is nonrelativistic ($v \ll 1$), disturbances 
tend to ``propagate'' superluminally, at $1/v$.  Hence, the NFA here is not 
a normal nonrelativistic reduction.  The resulting equations are 
``anti-Galilean'' invariant.  Consider two Lorentz frames, with the primed 
frame moving at velocity $w$ with respect to the unprimed frame.  
If $w$ is small, $O(\epsilon)$, then the Lorentz transformations become 
approximately ($\gamma_w \equiv 1/\sqrt{1-w^2}$)
\BE
\frac{\partial}{\partial t'} = \gamma_w \left( \frac{\partial}{\partial t} 
+ w \frac{\partial}{\partial x} \right) \approx  \frac{\partial}{\partial t}.
\EE
\BE
\frac{\partial}{\partial x'} = \gamma_w \left( \frac{\partial}{\partial x} 
+ w \frac{\partial}{\partial t} \right) \approx   \frac{\partial}{\partial x} 
+ w \frac{\partial}{\partial t}.
\EE
This is backwards with respect to the usual Galilean transformation, where the 
$x$ derivative would not change and the $t$ derivative would change by a $w$ 
term.  

   This is certainly strange, and takes some getting used to, but one should 
simply view it as an approximation to the full Lorentz transformations, 
valid in the stated context.  One is used to dealing with small objects that 
move slowly, so that their density distributions vary rapidly in space, but 
slowly in time.  In the present case one is dealing with large objects, 
slowly varying in space, but relatively rapidly varying in time.  This is 
related to the fact that the Higgs vacuum, as a spontaneous Bose-Einstein 
condensate, has almost all its particles in the same quantum state.  
Small disturbances of this state involve vast numbers of particles, spread 
over long distances, all moving nearly in lockstep, so that the disturbance 
varies only slowly with position while the whole collective has the same, 
relatively rapid time dependence.

%

\section{Shock speeds}

     Returning to the fundamental, relativistic energy and momentum 
conservation equations one can write down the two equations governing shock 
discontinuities.   In the empty-vacuum case one finds
\BE
\label{shockEV}
{\mbox{{\rm shock speed}}} = \frac{[[\rho(1+ g \rho) \gamma^2 v]]}
{[[(\rho(1 + g \rho) \gamma^2 - \half g \rho^2 ]]} 
= \frac{[[\rho(1 + g \rho )\gamma^2 v^2 + \half g \rho^2 ]]}
{[[\rho(1 + g \rho) \gamma^2 v]]},
\EE
which in the NFA indeed reduces to the usual, nonrelativistic result 
(\ref{shock}) with $P(\rho)=\half g \rho^2$.  

    In the Higgs-vacuum case the corresponding equations are 
\BE
\label{shockHV}
{\mbox{{\rm shock speed}}} = \frac{[[\sigma (1+ C \sigma) \gamma^2 v]]}
{[[\sigma(1 + C \sigma) \gamma^2 - \sigma(1+\half C \sigma)]]} 
= \frac{[[\sigma(1 + C \sigma )\gamma^2 v^2 + \sigma(1+\half C \sigma) ]]}
{[[\sigma(1 + C \sigma) \gamma^2 v]]}.
\EE
Since $\gamma^2 v^2 = \gamma^2-1$, one may re-write these equations as
\BE
\label{shockHVp}
{\mbox{{\rm shock speed}}} = \frac{[[\sigma (1+ C \sigma) \gamma^2 v]]}
{[[\sigma(1 + C \sigma) \gamma^2 v^2 + \half C \sigma^2]]} 
= \frac{[[\sigma(1 + C \sigma )\gamma^2 - \half C \sigma^2 ]]}
{[[\sigma(1 + C \sigma) \gamma^2 v]]}.
\EE
Comparing with (\ref{shockEV}) one finds that the shock speeds in the two 
cases are reciprocals of each other, in keeping with the $x \leftrightarrow t$ 
symmetry.  In the NFA the equations reduce to 
\BE
\label{shockHVpp}
{\mbox{{\rm shock speed}}} = 
\frac{[[\sigma v]]}{[[\sigma v^2 +\half C \sigma^2]]} 
= \frac{[[\sigma]]}{[[\sigma v]]}.
\EE

%

\section{An exact solution}

    In $1+1$ dimensions, the empty-vacuum hydrodynamic equations in the NFA, 
(\ref{ev1}, \ref{ev2}), are the well-known ``shallow-water-wave equations,'' 
a special case of the polytropic gas.  Thus, a general, exact 
$1+1$-dimensional solution can, in principle, be obtained using the method 
described in Sect. 98 of Ref. \cite{landau}.  I shall not pursue that program 
here, but rather I shall discuss a specific $1+1$ dimensional solution that 
has a clear physical interpretation in terms of the free expansion of an 
atomic Bose-Einstein condensate --- a phenomenon has been studied both 
theoretically \cite{dalfovo} and experimentally \cite{expts}.  From this 
solution, by $x \leftrightarrow t$, one can then obtain a solution to the 
$1+1$-dimensional Higgs-vacuum hydrodynamic equations; a solution that can 
then be generalized to a spherically symmetric solution in $d+1$ dimensions.   

     Ref. \cite{dalfovo} has found an exact, analytic solution of 
the empty-vacuum hydrodynamic equations in arbitrary dimensions.  Let us 
consider that solution in $1 \! + \! 1$ dimensions \cite{fnte1d}:
\BE 
\label{stringsol}
\left. 
\begin{array}{c}
g \rho = \frac{{\cal N}^2}{4}\frac{1}{b^3} 
\left( b^2 - \frac{x^2}{\ell^2} \right),
\\
v= {\scriptstyle{{\cal N}}} \sqrt{\frac{b-1}{b^3}} \frac{x}{\ell},
\end{array}
\right\}
\quad\quad -b \ell < x < b \ell,
\EE
where ${\scriptstyle{{\cal N}}}$ is a normalization constant and $b$ is a 
function of $t$ given by
\BE
\label{beq}
\sqrt{b(b-1)} + \ln(\sqrt{b} + \sqrt{b-1}) = 
\frac{{\scriptstyle{{\cal N}}}}{\ell} t.
\EE
It is straightforward to show that $b$ is a solution to the differential 
equations
\BE
\frac{db}{dt} = \frac{{\scriptstyle{{\cal N}}}}{\ell} \sqrt{\frac{b-1}{b}},
\EE
and 
\BE
\frac{d^2b}{dt^2} = \frac{1}{2 b^2} \frac{{\scriptstyle{{\cal N}}}^2}{\ell^2},
\EE
and hence to verify that Eqs. (\ref{ev1}, \ref{ev2}) are exactly satisfied. 
Note that the normalization constant ${\cal N}$ should be small and that 
the scaling of $v, g \rho$, $\partial/\partial t$ with ${\cal N}$ is 
then completely consistent with the requirements of the NFA, 
Eq. (\ref{NFAev}).  

      The function $b(t)$ starts from $b=1$ at $t=0$ and grows linearly for 
large $t$.  At $t=0$ the initial flow velocity is zero everywhere and the 
initial density distribution is an inverted parabola between $x=-\ell$ and 
$x=+\ell$ and zero outside.   The solution is self-similar, in that the 
density maintains an inverted parabolic form as its spreads out.  See Fig. 3.

      The initial density distribution is in fact the equilibrium density 
distribution in a harmonic trapping potential, in the Thomas-Fermi 
approximation.  Thus, the solution with these initial conditions is naturally 
realized in atom-trap experiments, simply by abruptly turning off a 
pre-existing trapping potential \cite{dalfovo, expts}.

\begin{figure}[htp]
\begin{center}
\includegraphics[width=8cm]{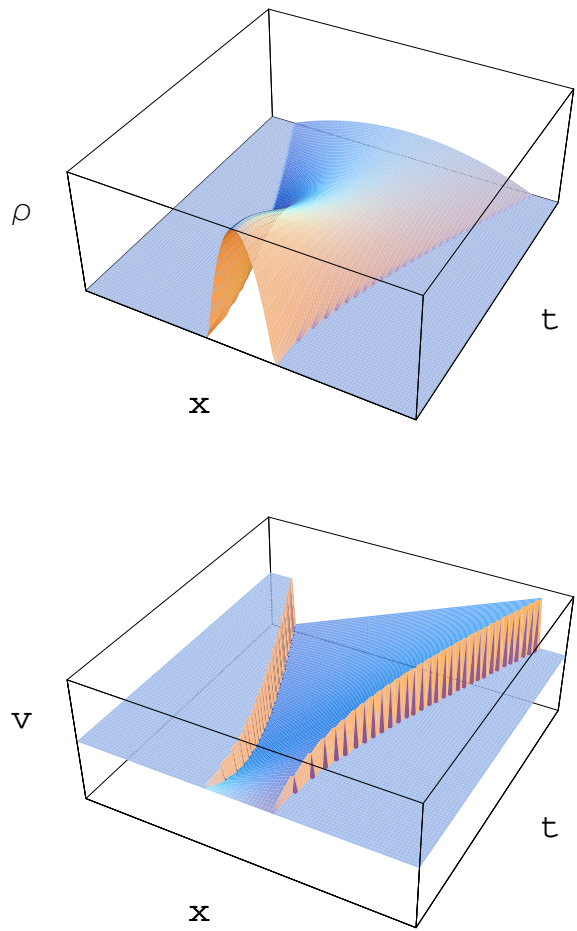}
\caption{Density and flow velocity as functions of $x$ and $t$ for the 
solution in Eq. (\ref{stringsol}).}
\end{center}
\end{figure}

      In Fig. 3. I have taken $v=0$ outside the range $-b \ell <x< +b \ell$. 
This choice is somewhat arbitrary because, of course, 
$v$ is physically meaningless anywhere where $\rho=0$.  The fundamental 
equations are mass and momentum conservation, (\ref{mass}) 
and (\ref{mtm}) with $P=\half g \rho^2$, which are identically satisfied 
if $\rho=0$, whatever $v$ may be doing.  The discontinuity in the 
derivative of $\rho$ at $x=\pm b \ell$, whether or not one wishes to regard 
it as a ``shock,'' is certainly a place where the hydrodynamic assumption 
of slow variation breaks down.  Corrections to hydrodynamics come into play 
to smooth out these ``corners'' in the density distribution \cite{pethick}.  
I discuss these corrections in the next section.

      By making the transformation $x \leftrightarrow t$ one can obtain an 
exact solution to the Higgs-vacuum hydrodynamic equations.  First note that 
the solution in Fig. 3 can be joined smoothly on to its time-reversed 
solution at negative times.  One can then take slices of Fig. 3 at various 
values of $x$, which becomes the new $t$.  This leads to Fig. 4 which shows 
the time evolution of a $\sigma$ disturbance initially concentrated around  
$x=0$, though with long tails out to infinity on each side, and with $v=0$ 
everywhere.  The disturbance spreads initially somewhat like the empty vacuum 
case, but then develops a dip at the centre; it then splits into two 
disturbances moving rapidly (superluminally) apart from each other, with 
each one spreading out.  Between them is a stretch of perfectly restored 
vacuum, $\sigma=0$.

\begin{figure}[htp]
\begin{center}
\includegraphics[width=11cm]{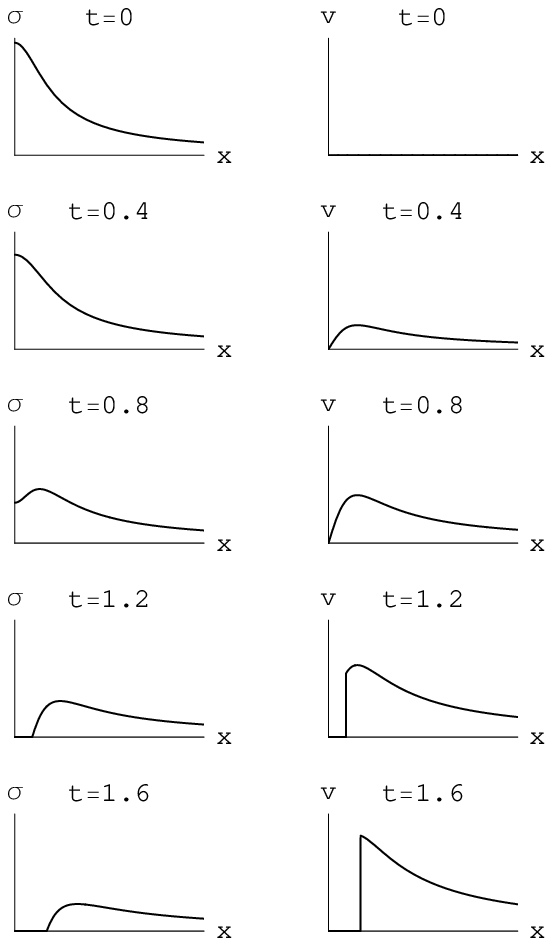}
\caption{Solution to the 1-dimensional Higgs-vacuum hydrodynamic equations 
obtained by $x \leftrightarrow t$ from Fig. 3. Only positive $x$ is shown; 
$\sigma$ is symmetric and $v$ is antisymmetric in $x$.  Time is in units of 
$\ell/{\cal N}$.  A qualitatively similar spherically symmetric solution 
exists in higher dimensions.}
\end{center}
\end{figure}

     This solution can be generalized to $d \! + \! 1$ dimensions.  The vacuum 
hydrodynamic equations (\ref{hv1}, \ref{hv2}) generalize to
\BE
\label{hvd1}
\delv \sigma + (\sigma \vec{{\bf{v}}})_t = 0,
\EE
\BE
\label{hvd2}
v v_t + \delv.\vec{{\bf{v}}} + \sigma_t = 0.
\EE
Assuming spherical symmetry, so that $\vec{{\bf{v}}}$ is radial, gives
\BE
\sigma_r + (\sigma v)_t = 0,
\EE
\BE
v_r + (d-1)\frac{v}{r} + (\half v^2 + C \sigma)_t = 0.
\EE
Note that the extra term, with respect to the $1 \! + \!1$ dimensional case, 
occurs in the second, not the first, equation.  One may now proceed to look 
for a self-similar solution of the form 
\BE 
\left. 
\begin{array}{c}
C \sigma = \frac{{\cal N}^2}{4}\frac{1}{b^3} 
\left( b^2 - \frac{t^2}{\tau^2} \right),
\\
v= {\scriptstyle{{\cal N}}} f \frac{t}{\tau},
\end{array}
\right\}
\quad\quad -b \tau < t < b \tau,
\EE
where $b$ and $f$ are functions only of 
$\hat{r} \equiv \frac{{\scriptstyle{{\cal N}}}}{\tau} r$.  Substituting this 
{\it Ansatz} in the first equation produces a solution if 
\BE
f =  \frac{b'}{b},
\EE
where $b' \equiv  \frac{db}{d\hat{r}}$. Then the second equation reduces to 
an ordinary differential equation for 
$b(\hat{r})$:
\BE
\label{bdiffeq}
b'' + (d-1) \frac{b'}{\hat{r}} - \frac{1}{2} \frac{1}{b^2} = 0. 
\EE
In the $1 \! + \! 1$ dimensional case this equation, multiplied through by 
$b'$, integrates up immediately to yield (\ref{beq}) with $t \to r$ and 
$\ell \to \tau$.  In higher dimensions the equation is less tractable, 
analytically.  However, it is easy to find a special, power-law solution 
valid for $d>\frac{4}{3}$:
\BE
b_0 = \left( 
\frac{3}{4} \frac{1}{(d-\frac{4}{3})}\right)^{1/3} \hat{r}^{2/3}.
\EE
With numerical methods one can find solutions that start from $b=1$ at 
$\hat{r}=0$ and approach this behaviour at large $\hat{r}$.  These 
solutions provide the analog to the $1 \! + \! 1$-dimensional solution 
discussed earlier, except that the large-$r$ dependence is $r^{2/3}$ rather 
than linear. The small-$\hat{r}$ behaviour is 
\BE 
b= 1+ \frac{1}{4 d} \hat{r}^2 + O(\hat{r}^4).
\EE
The subleading behaviour at large-$\hat{r}$ is rather curious.  Writing 
$b=b_0 + \beta$ and assuming $\beta \ll b_0$ yields a linear equation for 
$\beta$:
\BE
\beta'' + (d-1) \frac{\beta'}{\hat{r}} + 
\frac{4}{3} (d-\frac{4}{3}) \frac{\beta}{\hat{r}^2} = 0.
\EE
Putting $\hat{r}=e^z$ yields an equation with constant coefficients:
\BE
\ddot{\beta} + (d-2)\dot{\beta} + 
\frac{4}{3} (d-\frac{4}{3}) \beta = 0,
\EE
with solutions $\beta = e^{p z}$ with
\BE
p= \frac{1}{6} \left( 6 - 3 d \pm \sqrt{100-84 d + 9 d^2} \right).
\EE
For $d=8$ and above the roots are real, but for $d=2, \ldots 7$ the roots are 
complex.  For $d=3$ one finds 
\BE
\beta = \frac{c_1}{\sqrt{r}} 
\cos ({\scriptstyle{\frac{\sqrt{71}}{6}}} \ln r + c_2),
\EE
where $c_1,c_2$ are constants.  The presence of the oscillatory factor is, 
however, all but invisible in a plot of $b(\hat{r})$. See Fig. 5.  Since $b$ 
is qualitatively similar to the $1 \! + \! 1$ dimensional case (the main 
difference being the $r^{2/3}$ rather than linear growth at large $r$) the 
solution for $\sigma$ and $v$ as a function of radius at various times is 
qualitatively similar to Fig. 4.  

\begin{figure}[htp]
\begin{center}
\includegraphics[width=8cm]{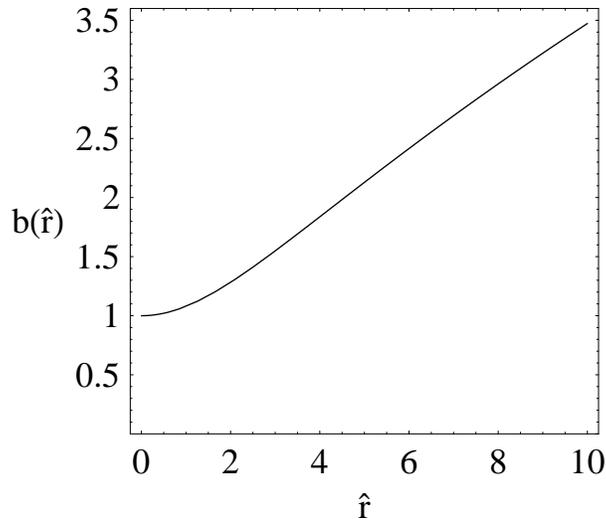}
\caption{The solution to Eq. (\ref{bdiffeq}) with $d=3$, for $b(0)=1$, $b'=0$. 
 ($\hat{r} = ({\cal N}/\tau) r$).}
\end{center}
\end{figure}

%

\section{Corrections to empty-vacuum hydrodynamics}

       In the atomic Bose-Einstein condensate literature the empty-vacuum 
hydrodynamic equations arise as an approximation to the Gross-Pitaevskii 
(GP) equation: \cite{pethick}
\BE 
\label{gpeq}
i \hbar \frac{\partial \psi}{\partial t} = -\frac{\hbar^2}{2 m} \delsq \psi 
+ V({\bf r}) \psi + U_0 \mid\! \psi \!\mid^2 \psi .
\EE
The constant $U_0=4 \pi a \hbar^2/m$, where $a$ is the $s$-wave scattering 
length is related to the constant $g$ introduced earlier by $g=U_0/m^2$.  
The GP equation itself is valid provided the medium is dilute; $n a^3 \ll 1$.  
In the present context there is no external potential, so $V({\bf r})=0$, and 
for simplicity I consider only the $1 \! + \! 1$ dimensional case.  

    Substituting
\BE
\psi = \sqrt{\frac{\rho}{m}} \, e^{i \varphi}, 
\quad\quad {\mbox{{\rm with}}} \quad 
v= \frac{\hbar}{m} \varphi_x,
\EE
and separating real and imaginary parts, one obtains two equations 
\cite{pethick,fntemad}.  One is mass conservation:
\BE
\label{gpsys1}
\rho_t + (\rho v)_x = 0,
\EE
and the other, after taking $\partial/\partial x$, becomes 
\BE
\label{gpsys2}
v_t + \left( \frac{1}{2} v^2 + g \rho - 
{\cal A} 
\left( \frac{\rho_{xx}}{\rho} - \frac{1}{2} \frac{\rho_x^2}{\rho^2} \right) 
\right)_x =0,
\EE
with ${\cal A}=\frac{1}{4} \frac{\hbar^2}{m^2}$.  These two equations reduce 
to the hydrodynamic ones when the density is so slowly varying that one may 
neglect the ``quantum pressure'' term proportional to ${\cal A}$.  

     Note that the corrections to empty-vacuum hydrodynamics involve  
{\it third} derivatives, rather than the second derivative terms 
characteristic of viscosity.  The corrections are non-dissipative and 
preserve the time-reversal ($t \to -t$, $v \to -v$) and parity 
($x \to -x$, $v \to -v$) symmetries of the hydrodynamic equations.  

\begin{table}[!hbtp]
\begin{center}
\begin{tabular}{|l|l|}
\hline
{\bf Burgers} & {\bf Empty vacuum hydrodynamics} \\
$v_t+ \left( \half v^2 \right)_x =0$  &
$v_t + \left( \half v^2 + g \rho \right)_x =0$ \\
$\rho_t + \left( \rho v \right)_x =0$ &
$\rho_t + \left( \rho v \right)_x =0$ \\
\hline
{\bf Schr\"odinger} & {\bf Gross-Pitaevskii} \\
$v_t+ \left( \half v^2 -{\cal A} \left( \frac{\rho_{xx}}{\rho} - \frac{1}{2} 
\frac{\rho_x^2}{\rho^2} \right) \right)_x =0$  &
$v_t + \left( \half v^2 + g \rho 
-{\cal A} \left( \frac{\rho_{xx}}{\rho} - \frac{1}{2} 
\frac{\rho_x^2}{\rho^2} \right) \right)_x 
=0$ \\
$\rho_t + \left( \rho v \right)_x =0$  & 
$\rho_t + \left( \rho v \right)_x =0$ \\
\hline
\end{tabular}
\end{center}
\caption{Family of four equation systems related to empty-vacuum 
hydrodynamics.  ${\cal A} \equiv \hbar^2/(4 m^2)$.}
\end{table}

     The two real equations (\ref{gpsys1}) and (\ref{gpsys2}) are equivalent 
to the single complex GP equation.  It is perhaps helpful to 
think of this system of equations as the senior member of a family of four 
equation systems, the others corresponding special cases with $g$ or 
${\cal A}$ or both set to zero; see Table 1.  With 
$g \! = \! {\cal A} \! = \! 0$ we have the Burgers system, corresponding 
to a pressureless gas of free, classical particles.  
This system has been studied recently by Choquard \cite{choquard}, who points 
out the solution $\rho \propto v_x$ \cite{fnteburgsol}.  Inclusion of particle 
interactions (non-zero $g$) leads to the empty-vacuum hydrodynamic equations.  
Returning to free particles ($g=0$) but including the quantum term 
(non-zero ${\cal A}$) corresponds to the free Schr\"odinger equation, which 
is manifestly {\it linear} in its usual complex form.  With both $g$ and 
${\cal A}$ non-zero one has the nonlinear GP equation.  Appendix A discusses 
the infinite tower of conservation laws associated with these equation 
systems.

%

\section{Corrections to Higgs vacuum hydrodynamics}

     What are the appropriate higher-derivative corrections to the 
hydrodynamic equations in the Higgs-vacuum case?  To answer this question, 
in the absence of a microscopic theory, necessarily involves some guesswork.  
Nevertheless some simple physical considerations, together with a 
mathematical/aesthetic criterion appear to point to a unique choice.  

     The physical considerations from the analogy with the empty vacuum case 
are (i) one expects third-derivative, non-dissipative, time-reversal invariant 
terms; (ii) in the NFA these will be time derivatives because space 
derivatives count as $O(\epsilon)$, (iii) the new constant multiplying the 
new terms should count as $O(\epsilon^2)$.  However, there is a further 
physical consideration that suggests that the new terms should break the 
$x \leftrightarrow t$ relation to the empty vacuum case:  
{\it unlike the empty-vacuum case, $v$ is meaningful even when} $\sigma = 0$.  
That is because in the Higgs case there is a pre-existing density of 
particles: a disturbance that causes these particles to move non-uniformly 
will cost energy, even if it nowhere changes the density at some initial 
instant. 

     Guided by these considerations and a mathematical/aesthetic criterion 
that the new equations, in $1 \! + \! 1$ dimensions, should be an integrable 
system, I looked for equation systems that would generate a tower of 
conservation laws.  In this way I was led to the following equations:
\BE
\label{new1}
v_x + \left( \half v^2 + C \sigma \right)_t =0,
\EE
\BE
\label{new2}
\sigma_x + \left( \sigma v - {\cal R} v_{tt} \right)_t =0.
\EE
The second of these equations represents momentum conservation.  As discussed 
in Appendix A, these two equations imply other conservation laws.  I conjecture
that there are an infinite number.  The next of these, in its simplest form, is
\BE
(\sigma v)_x + 
\left( \sigma v^2 + \half C \sigma^2 + 
{\cal R}(\half v_t^2 - v v_{tt}) \right)_t 
=0.
\EE
Alternative forms can be obtained by adding a total $x$ derivative to 
the density and subtracting the corresponding $t$ derivative from the flux.    
In this way one can obtain an energy-conservation equation whose energy flux 
coincides with the momentum density:
\BE
\left( \sigma v - {\cal R} v_{tt} \right)_x + 
\left( \sigma v^2 + \half C \sigma^2 
- {\cal R}(\half v_t^2 + 2 v v_{tt} + C \sigma_{tt}) \right)_t 
=0.
\EE
 
    It appears that the constant ${\cal R}$ must be positive, so that 
the new $v_{ttt}$ term in (\ref{new2}) has the effect of smoothing out the 
discontinuity in $v$ and the corresponding ``corner'' in $\sigma$ in the 
solution shown in Fig. 4 \cite{fnteR}.  

     Now, in analogy with Table 1, one has a family of four equation systems;  
see Table 2.  The pattern is somewhat different from Table 1, where the 
mass-conservation equation $\rho_t + (\rho v)_x=0$ held in each case.  
Its $x \leftrightarrow t$ analog in Table 2 (which is {\it not} 
mass conservation, but momentum conservation) becomes modified in the 
higher-derivative cases in the bottom half of the table.

\begin{table}[!hbtp]
\begin{center}
\begin{tabular}{|l|l|}
\hline
{\bf RIF} & {\bf Higgs-vacuum hydrodynamics} \\
$v_x+ \left( \half v^2 \right)_t =0$  &
$v_x + \left( \half v^2 + C \sigma \right)_t =0$ \\
$\sigma_x + \left( \sigma v \right)_t =0$ &
$\sigma_x + \left( \sigma v \right)_t =0$ \\
\hline
{\bf Schr\"odinger analog} & {\bf GP analog} \\
$v_x+ \left( \half v^2 \right)_t =0$  &
$v_x + \left( \half v^2 + C \sigma \right)_t 
=0$ \\
$\sigma_x + \left( \sigma v - {\cal R} v_{tt} \right)_t =0$  & 
$\sigma_x + \left( \sigma v - {\cal R} v_{tt} \right)_t =0$ \\
\hline
\end{tabular}
\end{center}
\caption{Family of four equation-systems related to Higgs-vacuum 
hydrodynamics.}
\end{table}

     The simplest case (top left in Table 2) is denoted ``RIF'' for 
``relativistic incompressible fluid'' since it corresponds to zero 
compressibility, $C=0$.  Its Euler equation $v_x+ v v_t=0$ 
is $\partial_\mu u^\mu =0$, the relativistic version of the 
usual incompressible-fluid condition $\delv.\vec{{\bf v}}=0$ \cite{landau}. 
The higher derivative cases, for lack of better names, are denoted as 
``Schr\"odinger analog'' and ``GP analog,'' even though the nature 
of the analogy is not clear. 

    I suspect that in the ``Schr\"odinger analog'' case there is some 
function $\Psi(\sigma,\varphi)$, with $\varphi_t=v$, that satisfies some 
{\it linear} equation, but I have been unable to find this function and 
equation.  I have found, though, that 
\BE
\sigma= -\frac{{\cal R}}{2} \left( \frac{v_{tt}}{v_t} \right)_t 
\EE
yields a solution to the $\sigma$ equation of the ``Schr\"odinger analog'' 
system.  More generally one can add to $\sigma$ a term proportional to 
$v_t$, since this is a solution in the RIF case.  
(Cf. \cite{choquard,fnteburgsol}.) 

     Introducing a velocity potential, $\varphi$, with $\varphi_t=v$, in the 
``GP analog'' system allows one to solve the first equation for $\sigma$:
\BE
C \sigma= -(\varphi_x + \half \varphi_t^2).
\EE
One may then use this to eliminate $\sigma$ in the second equation to obtain
\BE
\varphi_{xx}+2 \varphi_t \varphi_{tx} + 
\varphi_{tt} \left( \frac{3}{2} \varphi_t^2+ \varphi_x \right) 
+ {\cal R} C \varphi_{tttt} = 0.
\EE
Note that only the product of ${\cal R}$ and $C$ appears in this equation.

%

\section{`Bright' and `dark' solitons}
 
    For the GP equation it is well known that in the negative $g$ case, 
corresponding to attractive particle interactions, there are ``bright 
soliton'' solutions where the density has a $\sech^2$ form \cite{pethick}.  
(Such solitons have been observed experimentally \cite{hulet}.)
Remarkably, there are similar solutions to the ``GP analog'' equations, 
(\ref{new1}, \ref{new2}) if ${\cal R} C$ is negative.  Although this 
``wrong sign'' case probably has no physical significance \cite{fnteR} these 
solutions are of some interest because (i) they give support to the 
conjecture of integrability, and (ii) they hint at a symmetry, more subtle 
than $x \leftrightarrow t$, between the GP and ``GP analog'' equations.

    To begin, one postulates a travelling-wave solution
\BE
\label{trwave}
v=v(x-a t), \quad\quad\quad \sigma=\sigma(x-a t),
\EE
where the constant $a$ represents the soliton's velocity.  Substituting in 
Eq. (\ref{new1}) and integrating with respect to the variable 
$s\equiv x-a t$ gives
\BE
\label{sltnsig}
v -a(\half v^2 + C \sigma) =\mbox{{\rm const.}} \equiv \frac{(1-\xi^2)}{2 a}.
\EE
The constant of integration has been written in this form for later 
convenience.  Note that when $\sigma \to 0$ the velocity $v$ tends to one or 
other of the constant values $v_{\pm} \equiv (1 \pm \xi)/a$.
Substituting Eq. (\ref{trwave}) into Eq. (\ref{new2}) one obtains 
\BE
\sigma(1-a v) = - {\cal R} a^3 v'',
\EE
where the constant of integration must be zero to have $v''=0$ when $\sigma=0$.
Substituting for $\sigma$ from the previous equation leads to
\BE
(\xi^2-(1-a v)^2)(1- a v) =  - 2 {\cal R} C a^5 v''.
\EE
Multiplying through by $a v'$ and integrating yields
\BE
\label{sltnveq0}
4 {\cal R} C a^6 v'^2 = a v(2-a v)(2(1-\xi^2)-2 a v + a^2 v^2) + 
\mbox{{\rm const.}}.
\EE
The constant of integration here must be $-(1-\xi^2)^2$ so that $v' \to 0$ 
when $v \to v_{\pm} = (1\pm \xi)/a$.  The result then simplifies to 
\BE
\label{sltnveq}
v'^2 = -\frac{1}{4 {\cal R} C a^6} \left( (1-a v)^2 - \xi^2 \right)^2 .
\EE
Since the left-hand side is positive, a solution of this type is possible 
only if ${\cal R} C$ is negative.  It is convenient to define
\BE
\kappa \equiv \sqrt{-4 {\cal R} C}.
\EE
Taking the square root of Eq. (\ref{sltnveq}) and integrating yields the 
solution 
\BE
\label{bsv}
v= \frac{1}{a} \left[ 1 \pm \xi 
\tanh \left( \frac{\xi}{\kappa a^2} (x-a t) \right) \right].
\EE
From Eq. (\ref{sltnsig}) one obtains the corresponding form of $\sigma$:
\BE
\label{bss}
\sigma= \frac{\xi^2}{2 C a^2} 
\sech^2 \left( \frac{\xi}{\kappa a^2} (x-a t) \right).
\EE

     The above solution is characterized by two parameters $\xi$ and $a$ and 
by the sign choice, $\pm$, in Eq. (\ref{bsv}).  Each requires a little 
discussion.  (i) Without loss of generality $\xi$ may be taken as positive.  
For $\xi=0$ the solution becomes trivial ($\sigma=0$ and $v=$~constant, 
representing the vacuum itself).  For a single soliton one could set 
$\xi=1$ by choosing a frame of reference in which $v_{-}=0$.  However, for the 
multi-soliton case, discussed below, the extra generality of the $\xi$ 
parameter is important.  (ii) The $a$ parameter can have either sign but, for 
the NFA to be valid, the flow velocity must be small everywhere and hence 
$1/\!\mid \! a \! \mid \, \ll 1$.  Hence, the soliton moves at a hugely 
superluminal speed $\mid \! a \! \mid \gg 1$.  The total momentum and energy 
carried by the soliton are, however, small and {\it inversely} proportional 
to its speed and speed-squared, respectively:
\BE
P_{\rm soliton} = {\cal M} \left( \frac{1}{a} \right),   \quad \quad 
E_{\rm soliton} =  {\cal M} \left( \frac{1}{a} \right)^2,
\EE
with 
\BE
{\cal M} \equiv \int_{-\infty}^{\infty} \! dx \, \sigma = \xi \frac{\kappa}{C}.
\EE 
These results follow from evaluating the momentum and energy densities, which 
turn out to have the same $\sech^2$ form as $\sigma$; that is, using 
Eqs. (\ref{bsv}, \ref{bss}), one finds that
\BE
\sigma v - {\cal R} v_{tt} = \sigma/a,
\EE
\BE 
\sigma v^2 + \half C \sigma^2 
- {\cal R}(\half v_t^2 + 2 v v_{tt} + C \sigma_{tt})  
= \sigma/a^2.
\EE
(iii) Another important property of the solution is that 
\BE
\label{ansatz}
\sigma = \mp \frac{\kappa}{2 C} v_t ,
\EE
with the upper/lower sign determined by the upper/lower sign in 
(\ref{bsv}).   I shall refer to the lower-sign choice as ``solitons'' and the 
upper-sign choice as ``anti-solitons.''  Solitons have positive $v_t$ while 
anti-solitons have negative $v_t$.  

    Multi-soliton solutions can be constructed by using the {\it ansatz} 
$\sigma = \frac{\kappa}{2 C} v_t$.  Both of the ``GP analog'' equations are 
then satisfied if 
\BE
\label{vBur2}
v_x+ v v_t = -\frac{\kappa}{2} v_{tt},
\EE
which is the viscous Burgers' equation (\ref{vBurgers}) with 
$x \leftrightarrow t$.  This equation can be solved by the Hopf-Cole 
transformation \cite{fnteHopf}.  One defines  
\BE
\psi = e^{\varphi/\kappa}, \quad\quad {\rm with} \quad \varphi_t=v ,
\EE
and requires $\psi$ to satisfy the {\it linear} equation
\BE
\label{psieq}
\psi_x= - \frac{\kappa}{2} \psi_{tt}
\EE
(the diffusion equation with $x \leftrightarrow t$).  The resulting $v$ will 
then satisfy (\ref{vBur2}). The single-soliton solution, 
(\ref{bsv}, \ref{bss}) with the lower-sign choice, satisfies the 
$\sigma = \frac{\kappa}{2 C} v_t$ ansatz and by integrating $v$ one can find 
its corresponding $\varphi$.  The appropriate constant of integration is a 
function of $x$ chosen so that $\psi=e^{\varphi/\kappa}$ satisfies 
Eq. (\ref{psieq}).  Hence, one obtains the $\psi$ function corresponding to a 
single bright soliton.  By superposition, one can then write down a general 
class of solutions:
\BE
\psi= \sum_n K_n \exp \left(\frac{t}{\kappa a_n} \right) 
\exp \left( - (x-x_n) \frac{(1+\xi_n^2)}{2 \kappa a_n^2} \right) 
\cosh \left( \frac{\xi_n}{\kappa a_n^2} (x-x_n-a_n t) \right),  
\EE
where $K_n$, $a_n$, $\xi_n$, and $x_n$ are arbitrary constants.  By 
differentiation one can then obtain $v$ and $\sigma$ as
\BE
v= \kappa \frac{\psi_t}{\psi}, \quad\quad \sigma= \frac{\kappa^2}{2 C} 
\left( \frac{\psi_t}{\psi} \right)_t.
\EE

 Multi-anti-soliton solutions can be constructed in the same way by 
reversing the sign of $\kappa$ in the above.  However, it is not clear if one 
can analytically construct solutions involving both solitons and 
anti-solitons.  

    As noted earlier, these `bright soliton' solutions are probably 
unphysical since they require ${\cal R} C < 0$ \cite{fnteR}.  In the 
physical case, ${\cal R} C > 0$, there are analogs of the `dark soliton' 
solutions to the positive-$g$ GP equation.  One may find these by the 
same method as before, but in Eq. (\ref{sltnveq0}) one now needs a different 
constant of integration.  
Choosing it to be $2 \xi^2-1$ yields a solution where $v$ 
tends asymptotically to $1/a$ at $\pm \infty$. The resulting solution is
\BE
v=\frac{1}{a} \left[ 1 \pm \sqrt{4 a^2 C \sigma_0} \,
\sech \left( \sqrt{\frac{\sigma_0}{a^2 {\cal R}}} (x-a t) \right) \right],
\EE
\BE
\sigma= \sigma_0 \left[ 1 - 2 \sech^2 \! \left( 
\sqrt{\frac{\sigma_0}{a^2 {\cal R}}}  (x-a t) 
\right) \right],
\EE
where $\sigma_0 \equiv \xi^2/(2 a^2 C)$ is an arbitrary constant.  In this 
solution both $v$ and $\sigma$ are constant at $\pm \infty$ but have a 
localized ``dip'' or ``bump'' that moves with velocity $a$.  The dip/bump 
in $v$ can be arbitrarily small, but $\sigma$ has a dip that always goes 
negative, with a minimum value $-\sigma_0$.  
(See footnote \cite{fnteneg} for remarks on negative $\sigma$.) 

    Since this `dark soliton' solution does not have $\sigma \to 0$ at 
$\pm \infty$ it does not represent a perturbation of the vacuum.  However, one 
can regard it as describing a localized disturbance in a larger, and much 
more slowly varying, perturbation of the vacuum.  Thus, a solution like that 
shown in Fig. 4, for example, could be ``decorated'' with one or more `dark 
soliton' dips.  Calculating the momentum and energy densities for the dark 
soliton one finds, after subtracting the background values associated with 
$\sigma=\sigma_0$ and $v=1/a$, that they are $(\sigma-\sigma_0)/a$ and 
$(\sigma-\sigma_0)/a^2$, respectively.

    Note that the above solution can always be boosted to another frame with 
an ``anti-Galilean'' transformation to obtain another solution.  In particular 
one can boost to a frame where $v \to 0$ asymptotically.  In that frame $v$ 
and $\sigma$ become functions of $t$ only.

%

\section{Discussion}

     There are two main points that I wish to emphasize:  (i) Hydrodynamics 
in the empty-vacuum case makes perfect sense and describes an experimentally 
observed phenomenon, the free expansion of an atomic Bose-Einstein 
condensate when the atom-trap potential is turned off.  (ii) Hydrodynamics 
in the Higgs-vacuum case gives very strange and exciting behaviour as a 
consequence of the fact that the speed of sound in the Higgs vacuum 
is formally {\it infinite}.  The Higgs vacuum is a medium that is both 
ultrarelativistic (pressure $\gg$ energy density) and ultra-quantum, being a 
Bose-Einstein condensate with almost all its particles in the same quantum 
state.  Not surprisingly, perhaps, its properties are very different from 
those of familiar media.  

      The vastly superluminal speed of hydrodynamic disturbances in the Higgs 
case will obviously require careful study.  It may be important to point out 
that this phenomenon can occur without any material particle moving 
faster than the speed of light; see Fig. 6.  In fact, a similar issue arises 
even for a non-relativistic Bose-Einstein condensate \cite{fastsound}: 
phonons with low momentum, $k \to 0$, travel at a finite speed of sound, yet 
they are made up of atoms that move with very small velocity $k/m$.  In the 
non-relativistic case a spacetime picture like Fig. 6 predicts that the 
number of excited atoms present at any one time equals the ratio of the 
speed of sound to the speed of the individual atoms, a result that indeed 
follows from the usual Bogoliubov theory.  See Ref. \cite{fastsound} for a 
detailed discussion of this point.   

\begin{figure}[htp]
\begin{center}
\includegraphics[angle=90,width=9.5cm]{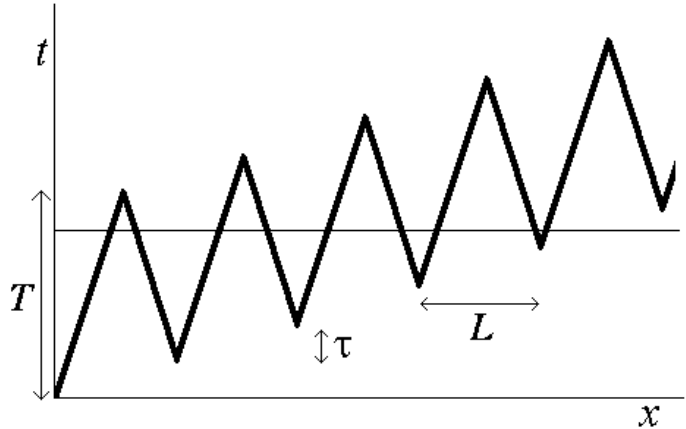}
\caption{Possible spacetime picture of a superluminal phonon-like excitation.  
There are multiple pair-creation and pair-annihilation events organized so 
that the collective motion is fast $L/\tau \gg c$ while the individual 
particles move slowly ($L/(2 T)<c$).}
\end{center}
\end{figure}

      There is an apparent dichotomy between hydrodynamics, which implies 
``soft'' modes that can be excited with arbitrarily little energy, and quantum 
field theory which (unless it explicitly contains massless particles) predicts 
an energy gap.  In the empty-vacuum case the atoms carry a conserved quantum 
number, so there is a superselection rule fixing the total number of atoms.  
The theory splits into separate, non-communicating ``sectors,'' each with 
a different total number of atoms.  The vacuum, strictly speaking, is in the 
zero-atom sector and its lowest excitation, an atom/anti-atom pair, requires 
an energy $2Mc^2$, where $M$ is the mass of the atom.  However, if one is 
given a system with $N$ atoms then the relevant ``vacuum'' state --- the 
lowest energy state with $N$ atoms --- has those atoms existing but dispersed 
in all directions to infinity (assuming a repulsive interaction $g>0$).  
This ``vacuum'' can be perturbed by bringing some or all of these atoms to 
finite distances from each other, and this may be done with arbitrarily little 
energy.  In this sense, there are soft, hydrodynamic modes.

     The situation for the Higgs vacuum is different.  No conserved quantum 
number is carried by the particles that spontaneously condense (the quanta 
of the scalar field that acquires a vacuum expectation value).  According to 
conventional theory, the only excitations are massive (quasi)particles, and 
hence the system has a gap.  However, the equation determining the inverse 
propagator at zero momentum actually has two solutions; $M_h^2$, where 
$M_h$ is the Higgs mass, and zero \cite{consoliprd}.  This fact suggests that 
there may be some deeper, underlying theory that possesses gap-less, 
hydrodynamic modes.  In the limit where this theory reduces to the 
conventional field theory, the realm of these hydrodynamic modes would shrink
to $p^\mu=0$.  Studying the hydrodynamic equations is then, I believe, a 
window into this unknown, deeper theory.  

     An example of what I have in mind is provided by the sine-Gordon 
equation \cite{SG} 
\BE 
\frac{\partial^2 \theta}{\partial t^2} - 
\frac{\partial^2 \theta}{\partial x^2} + \sin \theta =0.
\EE
For small $\theta$ this equation reduces to the Klein-Gordon equation and 
hence the dispersion relation is $\omega(k) = \sqrt{1+k^2}$.  One might 
therefore think that the spectrum of excitations necessarily has a gap.  
However, the non-linear equation possesses ``breather'' solutions
\BE
\theta(x,t) = 4 \arctan \left( \frac{\epsilon \sin(t/\sqrt{1+\epsilon^2})}
{\cosh (\epsilon (x-x_0)/\sqrt{1+\epsilon^2} \, )} \right),
\EE
whose fundamental frequency, $\omega_b=1/\sqrt{1+\epsilon^2} < 1$ lies in the 
gap.  (The sine-Gordon ``breather'' is unique in being stable.  Similar 
breathers exist in the $\phi^4$ case but they decay due to $e^{-1/\epsilon}$ 
couplings to linear, plane-wave modes \cite{ph4b}.)  The moral is that even 
if the linearized theory has only massive excitations, one cannot 
discount the possibility of soft, non-linear modes. 

      In this paper only the classical hydrodynamical equations have been 
considered.  A task for the future is to re-cast the equations into 
Hamiltonian form so that one can apply canonical quantization.  For a normal 
medium in the linear, acoustic regime this procedure would take one from 
classical sound waves to {\it phonon} quanta.  For the vacuum case the 
quantization will be inherently non-linear.  

\vspace*{7mm}

{\bf Acknowledgements}
I thank Maurizio Consoli for discussions.  This work was supported in part 
by the U.S. Department of Energy under Grant No. DE-FG05-97ER41031.

\newpage

%
%

\section*{Appendix A:  Towers of conservation laws}

     For each of the $1+1$ dimensional equation systems in Table 1 there is 
an infinite tower of conservation laws
\BE
D^{(n)}_t + F^{(n)}_x =0,
\EE
where $D$ is a density and $F$ is the associated flux.  Each $D^{(n)}$ will 
be  normalized such that its $\rho v^{n-1}$ term has unit coefficient.  
For $n=1,2,3$ the conserved quantities ($\int \! dx D^{(n)}$) are mass, 
momentum, and twice the nonrelativistic energy. Of course, one has the freedom 
to add a total $x$-derivative to a density; {\it i.e.}, $D'=D+\Gamma_x$ and 
$F'=F-\Gamma_t$ satisfy ${D'}_t+{F'}_x=0$.  

     The Burgers case is trivial, since it corresponds to non-interacting 
classical particles.  One has simply 
\BE
D^{(n)} = \rho v^{n-1}, \quad\quad\quad F^{(n)} = \rho v^n.
\EE
That is, not only are mass, momentum, and (nonrelativistic) energy 
(integrals of $\rho, \rho v,$ and $\half \rho v^2$) conserved, but --- since 
$v$ is constant for a free particle --- so is the integral of $\rho f(v)$ 
for any function of $v$.  (Thus, the set of conservation laws above is really 
only a subset of a continuous infinity of conservation laws. I focus on 
this subset because it generalizes to the other equation systems.)

    The empty-vacuum hydrodynamic equations include the effects of particle 
interactions (repulsive for $g>0$).  The $D$'s and $F$'s are now polynomials 
in $\rho$ and $v$:
\BE
D^{(n)} = v^{n+1}d_n(g \rho/v^2)/g, \quad\quad\quad 
F^{(n)} = v^{n+2}f_n(g \rho/v^2)/g
\EE
where the functions $d_n(s)$ and $f_n(s)$ (with $s=g \rho/v^2$) satisfy
\BE
(n+1) d_n = f_n' - (1-2 s) d_n',
\EE
\BE
(n+2) f_n=(1+2 s) f_n' -(1-s) d_n',
\EE
whose polynomial solutions are 
\BE
d_n(s) = \sum_{i=1}^{[(n+1)/2]} 
\frac{(n-1)!}{(n-2i+1)!} \frac{1}{i!(i-1)!} s^i,
\EE
\BE
f_n(s) =  \sum_{i=1}^{[(n+2)/2]} 
(n+1-i)\frac{(n-1)!}{(n-2i+2)!} \frac{1}{i!(i-1)!} s^i.
\EE
Two noteworthy relations are 
\BE
D^{(n)}= \frac{1}{n} \frac{\partial D^{(n+1)}}{\partial v},
\EE
\BE
F^{(n)}= \half (D^{(n+1)} + v D^{(n)}).
\EE

     In the Schr\"odinger case one goes back to free particles, but now they 
are quantum mechanical.  The $D$'s and $F$'s now involve higher derivatives. 
They can be expressed compactly in terms of the wavefunction $\psi$.  
Defining $\psi_{[N]}$ as the $N$th derivative $\psi_{xx \ldots x}$ one has
\BE
D^{(2N+1)}= \psi^*_{[N]} \psi_{[N]},
\EE
\BE
F^{(2N+1)}=D^{(2N+2)}= \frac{-i}{2} 
\left( \psi^*_{[N]} \psi_{[N+1]} - \psi^*_{[N+1]} \psi_{[N]} \right),
\EE
\BE
F^{(2N)}= \frac{1}{4} \left( 2 \psi^*_{[N]} \psi_{[N]} - 
\psi^*_{[N-1]} \psi_{[N+1]} - \psi^*_{[N+1]} \psi_{[N-1]} \right).
\EE
By substituting $\psi= \sqrt{\rho/m} \, e^{i \varphi}$ with 
$v=\frac{\hbar}{m} \varphi_x$ one can obtain the $D$'s and $F$'s in terms of 
$\rho$, $v$, and their derivatives, though the expressions quickly become 
cumbersome.  (See Table 3 with $g=0$ for the first few.)  There is a 
straightforward generalization of these results to $d\!+\!1$ dimensions, 
with the  $D^{(n)}$'s being alternately scalar and vector densities and the 
$F^{(n)}$'s correspondingly vector and tensor fluxes.   I have not seen these 
conservation laws for the free Schr\"odinger equation in any quantum-mechanics 
textbook, although they surely must be known.

     In the Gross-Pitaevskii (GP) case one has both quantum effects and 
interactions.  It is a highly non-trivial property of this equation, related 
to its integrability, that it admits an infinite number of conservation laws 
in the $1+1$ dimensional case \cite{zakharov}.  The first four of these are 
given explicitly in Table 3. 

\begin{table}[!hbtp]
\begin{center}
\begin{tabular}{|l|l|l|}
\hline
$n$ & $D^{(n)}$ & $F^{(n)}$ \\
\hline
1 & $\rho$ & $\rho v$ \\
\hline
2 & $\rho v$ & $\rho v^2 + \half g \rho^2 - 
{\cal A} \left( \rho_{xx} - \frac{\rho_x^2}{\rho} \right)$ \\
\hline
3 & $\rho v^2 + g \rho^2 + {\cal A} \frac{\rho_x^2}{\rho}$ & 
$\rho v^3 + 2 g \rho^2 v + {\cal A} \left( 
\frac{3 \rho_x^2 v}{\rho} + 2 \rho_x v_x - 2 \rho_{xx} v \right)$ \\
\hline
4 & $\rho v^3 + 3 g \rho^2 v$  & 
$\rho v^4 + \frac{9}{2} g \rho^2 v^2 + g^2 \rho^3$ \\
 & ${\mbox{}} + {\cal A} \left( 
\frac{3 \rho_x^2 v}{\rho} + 2 \rho_x v_x - 2 \rho_{xx} v \right)$ & 
${\mbox{}} + 
{\cal A} \left(-5 \rho_{xx} v^2 + 6 \frac{\rho_x^2}{\rho} v^2 + 4 \rho_x v v_x 
+ 2 \rho v_x^2 - 2 \rho v v_{xx} \right)$ \\
 & & ${\cal A} g \left( \frac{9}{2} \rho_x^2 - 3 \rho \rho_{xx} \right)$ \\
 & & ${\mbox{}} + 
{\cal A}^2 \left( \frac{-2 \rho_{xxx} \rho_x}{\rho} + 
\frac{2 \rho_{xx}^2}{\rho} + 
\frac{\rho_{xx} \rho_x^2}{\rho^2} - \frac{\rho_x^4}{\rho^3} \right)$ \\
\hline
\end{tabular}
\end{center}
\caption{The first four conservation laws, $D_t+F_x=0$, for the GP equation 
system. ${\cal A} = \hbar^2/(4 m^2)$. $D^{(1)}$ is the mass density, 
$D^{(2)}$ is the momentum density, and $D^{(3)}$ is twice the nonrelativistic 
energy density. }
\end{table}

     I turn now to the equation systems of Table 2.  The RIF and Higgs-vacuum 
hydrodynamics cases are related by $x\leftrightarrow t$ to the Burgers and 
empty-vacuum cases, respectively and so inherit the same set of conservation 
laws as 
\BE
D^{(n)}_x + F^{(n)}_t =0. 
\EE
(Note that I choose {\it not} to switch the names $D$ and $F$, so one needs 
to remember that now $D$ is a flux and $F$ is a density.) 

     For the ``Schr\"odinger analog'' case (${\cal R} \neq 0$, but $C=0$) it 
is easily verified that there are an infinite number of conservation laws 
that need only involve $\sigma, v, v_t, v_{tt}$:
\BE
F_n = \sigma v^n 
+ \half {\cal R} \left( (n-1)^2 v^{n-2} v_t^2 -2 v^{n-1} v_{tt} \right),
\EE
\BE
D_n = \sigma v^{n-1} 
+ \half {\cal R} (n-1)(n-2) v^{n-3} v_t^2
\EE
(One may of course modify these by adding a total $t$-derivative to $D$ 
and subtracting the corresponding $x$-derivative from $F$.  I have chosen 
to do this for the $n=2$ case in Table 4 in order for $D^{(2)}$, interpreted 
as the energy flux, to equal the momentum density $F^{(1)}$.) 

      Table 4 shows the first four conservation laws for the ``GP analog'' 
system.

\begin{table}[!hbtp]
\begin{center}
\begin{tabular}{|l|l|l|}
\hline
$n$ & $D^{(n)}$ & $F^{(n)}$ \\
\hline
1 & $\sigma$ & $\sigma v - {\cal R} v_{tt}$ \\
\hline
2 & $\sigma v-{\cal R} v_{tt}$ & $\sigma v^2 + \half C \sigma^2 - 
{\cal R} (\half v_t^2 + 2 v v_{tt} + C \sigma_{tt})$ \\
\hline
3 & $\sigma v^2 + C \sigma^2 + {\cal R} v_t^2$  & 
$\sigma v^3 + 2 C \sigma^2 v + {\cal R} (2 v v_t^2 - v^2 v_{tt} + 
2 C ( \sigma_t v_t - \sigma v_{tt}))$ \\
\hline
4 & $\sigma v^3 + 3 C \sigma^2 v$  & 
$\sigma v^4 + \frac{9}{2} C \sigma^2 v^2 + C^2 \sigma^3 
+ {\cal R} (\frac{9}{2} v^2 v_t^2 - v^3 v_{tt}) $ \\
 & ${\mbox{}} + {\cal R}(3 v v_t^2 - 4 C \sigma v_{tt})$ & 
${\mbox{}} + {\cal R} C (6 \sigma_t v v_t - 3 \sigma v_t^2 
- 10 \sigma v v_{tt})  $ \\
 & & ${\mbox{}} + 2 {\cal R}^2 C v_{tt}^2 + 
{\cal R} C^2 (2 \sigma_t^2 - 4 \sigma \sigma_{tt})$ \\
\hline
\end{tabular}
\end{center}
\caption{The first four conservation laws, $D_x+F_t=0$, for the ``GP analog'' 
system (Higgs-vacuum hydrodynamics plus higher-derivative corrections).  
$F^{(1)}$ is the momentum density and $F^{(2)}$ is the energy density. }
\end{table}

%
%


\begin{thebibliography} {99}
\bibitem{fnteg}
$g$ may be negative, but then higher-order terms in ${\cal E}$ are crucial 
to ultimate stability.
\bibitem{fnteslow}
Note that although the flow is nonrelativistic, $v \ll 1$, it is not 
``slow'' because $v$ is always much greater than the speed of sound, which 
is zero here.
\bibitem{fnteburgsol}
The $\rho$ equation in the Burgers' case admits a more general solution, 
$\rho= f(v) v_x + g(v) (v^2 - 2 v_x \varphi )$,  
where $f$ and $g$ are arbitrary functions and $\varphi_x=v$.  However, only 
the case $f=$~constant and $g=0$ is consistent with Galilean invariance.  
In the relativistic case one would need $f \propto \gamma = 1/\sqrt{1-v^2}$ 
for Lorentz invariance.
\bibitem{fnteR}
However, it is conceivable that ${\cal R}$ may be negative in the same sense 
that $g$ can be negative in the empty-vacuum case; see footnote \cite{fnteg}.
\end{thebibliography}
\end{document}